\documentclass[reprint,aps,prd]{revtex4-1}

\usepackage{amsmath}
\usepackage{xcolor}
\usepackage{graphicx}
\usepackage{siunitx}
\usepackage[
    pdfauthor={Sebastian Steinlechner},
    pdftitle={Local-Oscillator Noise Coupling in Balanced Homodyne Readout for Advanced Gravitational Wave Detectors},
    colorlinks=true,
    linkcolor=black,
    urlcolor=blue,
    citecolor=blue
]{hyperref}

\DeclareMathOperator{\var}{\Delta^2\!}

\newcommand{\RIN}{\mathrm{RIN}}
\newcommand{\op}[1]{\boldsymbol{#1}}

\newcommand{\opd}[1]{\op{#1}^{\dagger}}
\newcommand{\dop}[1]{\delta\mkern-1mu\op{#1}}

\newcommand{\re}{\mathrm{e}}
\newcommand{\ri}{\mathrm{i}}

\newcommand{\eps}{\varepsilon}
\newcommand{\figref}[1]{Fig.~\ref{fig:#1}}

\DeclareSIUnit{\sqrthz}{\ensuremath{\sqrt{\text{\hertz}}}}

\begin{document}
\title{Local-Oscillator Noise Coupling in Balanced Homodyne Readout for Advanced Gravitational Wave Detectors}

\author{Sebastian Steinlechner}
\author{Bryan W Barr}
\author{Angus S Bell}
\author{Stefan L Danilishin}
\author{Andreas Gl\"afke}
\author{Christian Gr\"af}
\author{Jan-Simon Hennig}
\author{E Alasdair Houston}
\author{Sabina H Huttner}
\author{Sean S Leavey}
\author{Daniela Pascucci}
\author{Borja Sorazu}
\author{Andrew Spencer}
\author{Kenneth A Strain}
\author{Jennifer Wright}
\author{Stefan Hild}
\affiliation{SUPA, School of Physics and Astronomy, The University 
of Glasgow, Glasgow, G12\,8QQ, UK}

\begin{abstract}
The second generation of interferometric gravitational wave detectors are quickly approaching their design sensitivity. For the first time these detectors will become limited by quantum back-action noise. Several back-action evasion techniques have been proposed to further increase the detector sensitivity. Since most proposals rely on a flexible readout of the full amplitude- and phase-quadrature space of the output light field, balanced homodyne detection is generally expected to replace the currently used DC readout. Up to now, little investigation has been undertaken into how balanced homodyne detection can be successfully transferred from its ubiquitous application in table-top quantum optics experiments to large-scale interferometers with suspended optics. Here we derive implementation requirements with respect to local oscillator noise couplings and highlight potential issues with the example of the Glasgow Sagnac Speed Meter experiment, as well as for a future upgrade to the Advanced LIGO detectors. 
\end{abstract}

\maketitle

%%% =========================================================================

\section{Introduction}
\label{sec:introduction}

A century after gravitational waves were first predicted by Albert Einstein as a result of his General Theory of Relativity \cite{Einstein18}, the worldwide effort to provide the first direct detection of these waves is still ongoing. The Advanced LIGO generation of interferometric gravitational wave detectors have just left their construction stages and are now being brought to their design sensitivity \cite{AdvLigo2015}.

Contrary to previous generations, these detectors will be limited by quantum noise over their whole detection bandwidth. Above about 50\,Hz, shot noise from the photon statistics will be dominant, while at lower frequencies the sensitivity will be limited by back-action noise. In anticipation of these limitations, several technologies have been devised over the past years to specifically target and reduce quantum noise. Squeezed states of light are the most mature of these technologies and their application in gravitational wave detectors has been successfully demonstrated in GEO\,600 and LIGO \cite{Abadie2011,Aasi2013} and intensely studied over the past few years \cite{Grote2013}.

To target the full quantum-noise spectrum, squeezing has to be combined with filter cavities that shape the quantum correlations such that an optimal back-action suppression is achieved at each frequency \cite{Kimble01}. Furthermore, topologies such as variational readout have been proposed where also the quadrature of the detected output light field is adjusted in a frequency dependent way \cite{Kimble01}. This is an example of a quantum non-demolition measurement, which theoretically allows for a back-action noise free readout \cite{Braginsky1980}. Another example for such a measurement is the speed meter configuration, where the velocity of the test masses is detected instead of their position \cite{Braginsky90}. It exploits that the momentum -- and the derived quantity velocity -- of a free test mass is conserved, and thus repeated measurements of the speed do not influence each other. In 2003, Y.~Chen showed that the Sagnac interferometer configuration is a close approximation of a speed meter \cite{Chen2003}, and experiments are now underway to demonstrate the back-action evasion properties of Sagnac interferometers \cite{Graef2014}. 

In these advanced technologies, it is generally assumed that the detected field quadrature of the output signal can be freely adjusted, which is essential for their ability to surpass the standard quantum limit. This is however impossible with the current DC readout technique which is only sensitive to the amplitude quadrature \cite{Hild2009,Fricke2012}. Balanced homodyne (BHD) detection \cite{*[{see e.g. }] [{}] Leonhardt1997} would allow for precisely choosing the readout quadrature \cite{DiGuglielmo2011}, and it is a well-proven workhorse for table-top experiments throughout quantum optics \cite{Bachor2004}. It has been shown to provide shot-noise limited sensitivity over a wide frequency band from the GHz range \cite{Ast2013} down to the sub-Hz regime \cite{Stefszky2012}. 

However, attention has only recently turned to the problem of how to implement BHD readout in large-scale interferometers \cite{Fritschel2014}. Here, we investigate the noise requirements that have to be imposed on such a readout scheme. In particular we focus on the case of significant amounts of carrier light (on the order of a \si{mW}) in the interferometer output field. This carrier light is introduced by a contrast defect of the main interferometer (e.g.\ differential losses of the interferometer arms) and, as we will show below, can lead to significant requirements on local oscillator stability.

%%% =========================================================================

\section{Fundamentals of balanced homodyne detection}

\begin{figure}
  \includegraphics{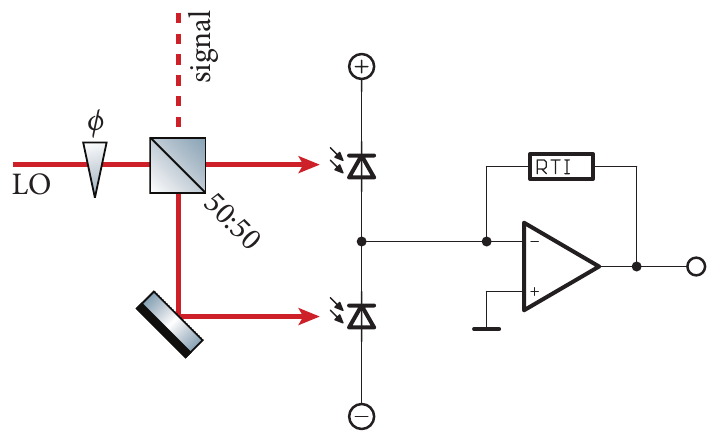}
  \caption{Schematic of balanced homodyne detection. The signal field is overlapped on a 50:50 beam splitter with a strong local oscillator (LO). Both beam splitter outputs are detected with a high-efficiency photo diode and subtracted from each other. Here, a direct photo-current subtraction circuit is shown. The readout quadrature can be selected by adjusting the relative phase $\phi$ between signal and LO field. RTI, transimpedance-gain setting resistor.}
  \label{fig:current_subtracting}
\end{figure}

For the following discussion, a simple review of the quantum-mechanical description of balanced homodyne detection is helpful. Following e.g. \cite{Gerry05}, we define the amplitude and phase quadrature operators $\op{X}_1 = (\op a+\opd a)/2$, $\op{X}_2 = -\ri(\op a-\opd a)/2$, where $\op a$ is the annihilation operator for a single-mode optical field. Furthermore, we set $\op a = \alpha + \delta\op{a}$, which separates the mode's classical amplitude $\alpha$ from its quantum fluctuations $\delta\op{a}$. In a step that is commonly referred to as linearisation, terms of higher than linear order in the noise are  neglected, i.e.\ $\delta\op{a}\delta\op{b}\to 0$.

In balanced homodyne detection, the signal beam $\op b$ is overlapped at a 50:50 beam splitter with a strong local oscillator field (LO) $\op a$ as shown in \figref{current_subtracting}. Without loss of generality, we can assume their coherent amplitudes $\alpha$ and $\beta$, respectively, to be real, and absorb the relative phase $\phi$ between the two fields in a phase factor $\exp(\ri \phi)$. The intensity in the two beam splitter output ports $\op c$ and $\op d$ is then given by
\begin{align}
    \opd c \op c &= \frac12 \Bigl[\opd a \op a + \opd a \op
        b \re^{-\ri\phi} + \op a \opd b \re^{\ri\phi} + \opd b \op b\Bigr] \\
      & = \frac12 \Bigl[\alpha^2 + \beta^2 + 2\alpha\beta\cos\phi \nonumber\\
      & \qquad + 2\alpha\dop{X}_1^a + 2\beta\dop{X}_1^b \nonumber\\
      & \qquad + 2\alpha\dop{X}_{-\phi}^b + 2\beta\dop{X}_\phi^a \Bigr]
\intertext{and}
    \opd d \op d &=\frac12 \Bigl[\alpha^2+\beta^2 - 2\alpha\beta\cos\phi \nonumber\\
      & \qquad + 2\alpha\dop{X}_1^a + 2\beta\dop{X}_1^b \nonumber\\
      & \qquad - 2\alpha\dop{X}_{-\phi}^b - 2\beta\dop{X}_\phi^a \Bigr]\,.
\end{align}
Here, $\op X_\phi = \op X_1 \cos\phi + \op X_2\sin\phi$ is the quadrature operator for the quadrature angle $\phi$. A more detailed calculation can be found in the appendix in Appendix~\ref{sec:detailed_calculations}.

Both output ports $\op c$ and $\op d$ are separately detected with high-efficiency photo detectors and the resulting photo currents are subtracted, yielding a detector output that is proportional to
\begin{align}
  \op{i_-} = \opd c \op c - \opd d\op d = 2\alpha\beta\cos\phi + 2\alpha\delta\op{X}_{-\phi}^b + 2\beta\delta\op{X}_{\phi}^a\,,
  \label{eq:bhd_diff_out}
\end{align}
with a noise variance of
\begin{align}
  \var \op{i_-} = 4\alpha^2\var{\delta\op X_{-\phi}^b} + 4\beta^2\var{\delta\op X_\phi^a}\,.
  \label{eq:bhd_diff_out_var}
\end{align}
The power in the local oscillator $P_{\rm LO} = \alpha^2$ is chosen to be much larger than the power in the signal field $P_{\rm sig} = \beta^2$,
\begin{align}
  P_{\rm LO} \gg P_{\rm sig}\,.
  \label{eq:lo_gg_sig}
\end{align}
Therefore the last term in Eq.~\eqref{eq:bhd_diff_out_var} can usually be neglected. The output is then directly proportional to the signal's noise in the quadrature $\op X_{-\phi}^b$, amplified by the coherent amplitude of the local oscillator. Tuning the relative phase $\phi$ between local oscillator and signal allows for an easily accessible adjustment of the detected quadrature.

Fulfilling the condition \eqref{eq:lo_gg_sig} is however not sufficient when the noise variance of the local oscillator beam is much higher than the signal's noise variance. Usually, the signals that one tries to measure with balanced homodyne detection are very close to or even below the quantum-mechanical zero-point fluctuations, i.e.\ $\var\op\delta X_{\phi}^b \approx \hbar\omega/4$. At the same time, the local oscillator will be orders of magnitude away from the shot-noise limit due to technical laser noise, unless the measurement frequencies are well above the laser's relaxation oscillation, or else significant effort has been put into laser stabilisation \cite{Kwee2009}. Therefore, even tiny amounts of carrier light in the signal field can amplify the local oscillator's noise sufficiently such that it completely dominates the output of the balanced homodyne detector. In the following section we will derive requirements on the amplitude and phase stability of the local oscillator, depending on the residual power $P_{\rm sig}$ in the signal beam.

%%% =========================================================================

\section{Noise coupling mechanisms in balanced-homodyne detection}

\textbf{Amplitude noise} -- 
Current DC readout schemes set strong requirements to the laser amplitude noise level: To provide the local oscillator light, the interferometer output port has to operate at some DC offset, thus sending a fraction of the light from the laser directly towards the output detector.

Balanced homodyne readout also uses a local oscillator at the carrier frequency and is thus in principle susceptible to laser amplitude noise. However, this noise is common to both photo diodes and can thus be subtracted out by careful balancing of the photo currents. We can see this behaviour by putting an imbalance term $1-\eps$ into \eqref{eq:bhd_diff_out}, see also Appendix~\ref{sec:cmrr} for more details:
\begin{align}
  \op i_{-,\eps} &= \opd c \op c - (1-\eps) \opd d\op d \\
    &= \text{const} + \eps\alpha\delta\op X_1^a + (2-\eps)\alpha\delta\op{X}_{-\phi}^b\,.
\end{align}
Any difference in e.g.\ beam-splitting  ratio and photo diode quantum efficiency thus adds noise from the local oscillator's amplitude quadrature $\op X_1^a$. In practice, $\eps$ can be made very small and common-mode suppression ratios of more than 80\,dB have been demonstrated \cite{McKenzie2007}.

The situation gets worse, however, when the signal field carries significant DC power, i.e.\ $\beta \gg 0$. Such a situation quickly arises due to a small contrast defect of the interferometer. Here we derive a relative intensity noise (RIN) requirement for the LO beam as follows. Looking at the noise variance in the output signal \eqref{eq:bhd_diff_out_var}, we see that the noise in the signal field is amplified by the LO carrier, while noise in the LO field is amplified by the signal carrier. We set $\phi = 0$ as we are only concerned with noise in the amplitude quadrature for now. The output should be dominated by the noise in the signal, therefore we require
\begin{align}
  4 P_{\rm LO} \var\delta\op X_1^b > 4P_{\rm sig}\var\delta\op X_1^a\,.
\end{align}
After dividing both sides by $P_{\rm LO}\times P_{\rm sig}$, i.e.\ the product of the mean (DC) powers in both signal and LO, this takes on the form of
\begin{align}
  \RIN^2_{\rm sig} > \RIN^2_{\rm LO}\,,
\end{align}
where
\begin{align}
  \RIN = \frac{\Delta P}{P} = \frac{\sqrt{P\var\delta\op X_1}}{P}
\end{align}
is the relative intensity noise of the respective beam. Finally, for signals at the quantum noise limit, $\RIN_{\rm sig}$ is just the shot-noise limited $\RIN$, $\RIN_{\rm SN, sig} = \sqrt{2\hbar\omega/P_{\rm sig}}$, and thus results in the requirement
\begin{align}
  \RIN_{\rm LO} < \sqrt{\frac{2\hbar\omega}{P_{\rm sig}}}\,.
  \label{eq:rin_requirement}
\end{align}
Since balanced homodyne detection requires $P_{\rm LO} > P_{\rm sig}$, this inequality is always fulfilled as long as the LO field is shot-noise limited. This is usually not a problem in table-top experiments in quantum optics where typical signals occur at sideband frequencies of several MHz, well away from technical noise sources such as the laser's relaxation oscillation. For signals in the audio-band, however, it is much more difficult to get close to a shot-noise limited $\RIN$ \cite{Kwee2009} so that the condition~\eqref{eq:rin_requirement} can pose a challenging requirement.

\begin{figure}
  \includegraphics[width=\linewidth]{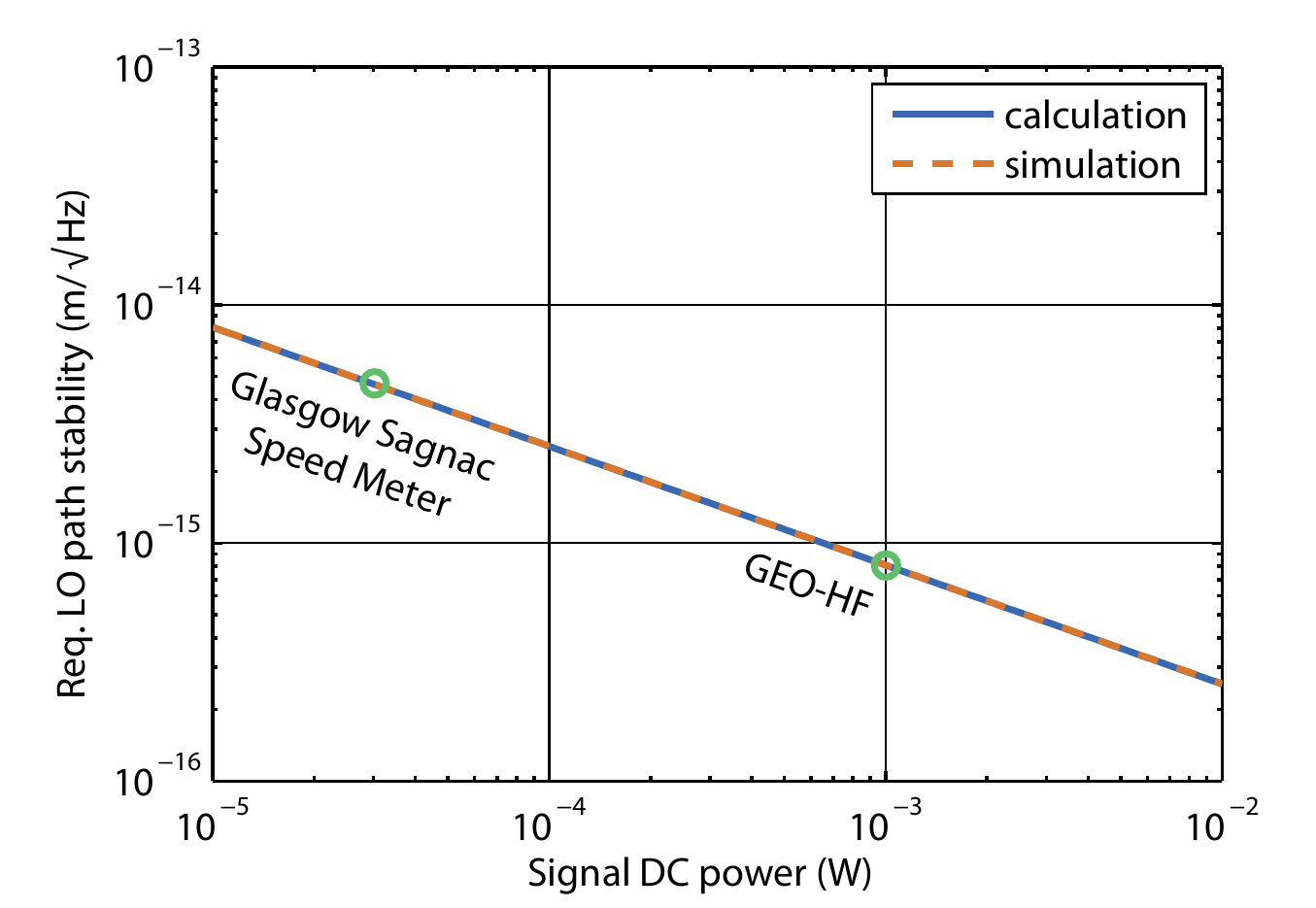}
  \caption{Requirement on local oscillator path length stability depending on the residual DC power level in the signal field. The blue curve is calculated from \eqref{eq:pn-req-x} for a wavelength of \SI{1064}{nm}, while the orange dashed curve is a simulated curve using the interferometry simulation tool Finesse \cite{*[{}] [{, the program is available at \url{http://www.gwoptics.org/finesse}.}] Freise2004}. We have indicated the requirements for the Glasgow Sagnac Speed Meter, see Section~\ref{sec:glasgow_ssm} below. In addition, we have added an indicative value for GEO-HF based on the residual fundamental mode content in its output port \cite{Grote2015}.}
  \label{fig:lo_path_stability}
\end{figure}

\textbf{Local oscillator path length stability} -- Let us now address phase noise within the local oscillator beam. In particular, since the LO and signal beams will have to travel along spatially separate paths for at least some distance, we will derive a requirement for the differential length fluctuations between these two paths.

Again, from Eq.~\eqref{eq:bhd_diff_out_var} we get the requirement that
\begin{equation}
    P_{\rm sig}\var\op X_2^{\rm LO} < P_{\rm LO} \var\op X_2^{\rm sig} \approx P_{\rm LO}\frac{\hbar\omega}{4} \,,
    \label{eq:pn-req-1}
\end{equation}
such that the BHD output is not limited by excess noise on the LO beam. In the last step, we assumed a signal that is close to the vacuum noise, i.e.\ $\var\op X_2^{\rm sig} \approx \hbar\omega/4$. If a (suspended) mirror in the LO path moves by a distance $\Delta x$, the corresponding signal in the phase quadrature is given by \cite{Danilishin2012}
\begin{equation}
    \op X_2^{\rm LO} = \frac{2\omega}{c}\sqrt{P_{\rm LO}}\Delta x\,.
\end{equation}
Inserting this into \eqref{eq:pn-req-1} leads to
\begin{equation}
    P_{\rm sig} \frac{4\omega^2 P_{\rm LO}}{c^2} \var x < P_{\rm LO}\frac{\hbar\omega}{4}
\end{equation}
and thus
\begin{equation}
    \var x < \frac{\hbar c^2}{16 P_{\rm sig}\omega}\,,
    \label{eq:pn-req-end}
\end{equation}
which is, perhaps surprisingly, independent of $P_{\rm LO}$. Equivalently, we get the requirement for the single-sided displacement spectral density
\begin{equation}
    \tilde x(f) < \sqrt{\frac{\hbar c^2}{8\omega P_{\rm sig}}}
        \approx \SI{8.2e-16}{m/\sqrthz} \times \sqrt{\frac{\SI{1}{mW}}{P_{\rm sig}}}
    \label{eq:pn-req-x}
\end{equation}
assuming a carrier frequency of $\omega = 2\pi c / \SI{1064}{nm}$.
This requirement is flat over the whole frequency range that is of interest for the measurement. It is shown as a function of $P_{\rm sig}$ in \figref{lo_path_stability}, together with the results of a Finesse \cite{Freise2004} simulation for this setup.

\begin{figure}
    \includegraphics[width=\linewidth]{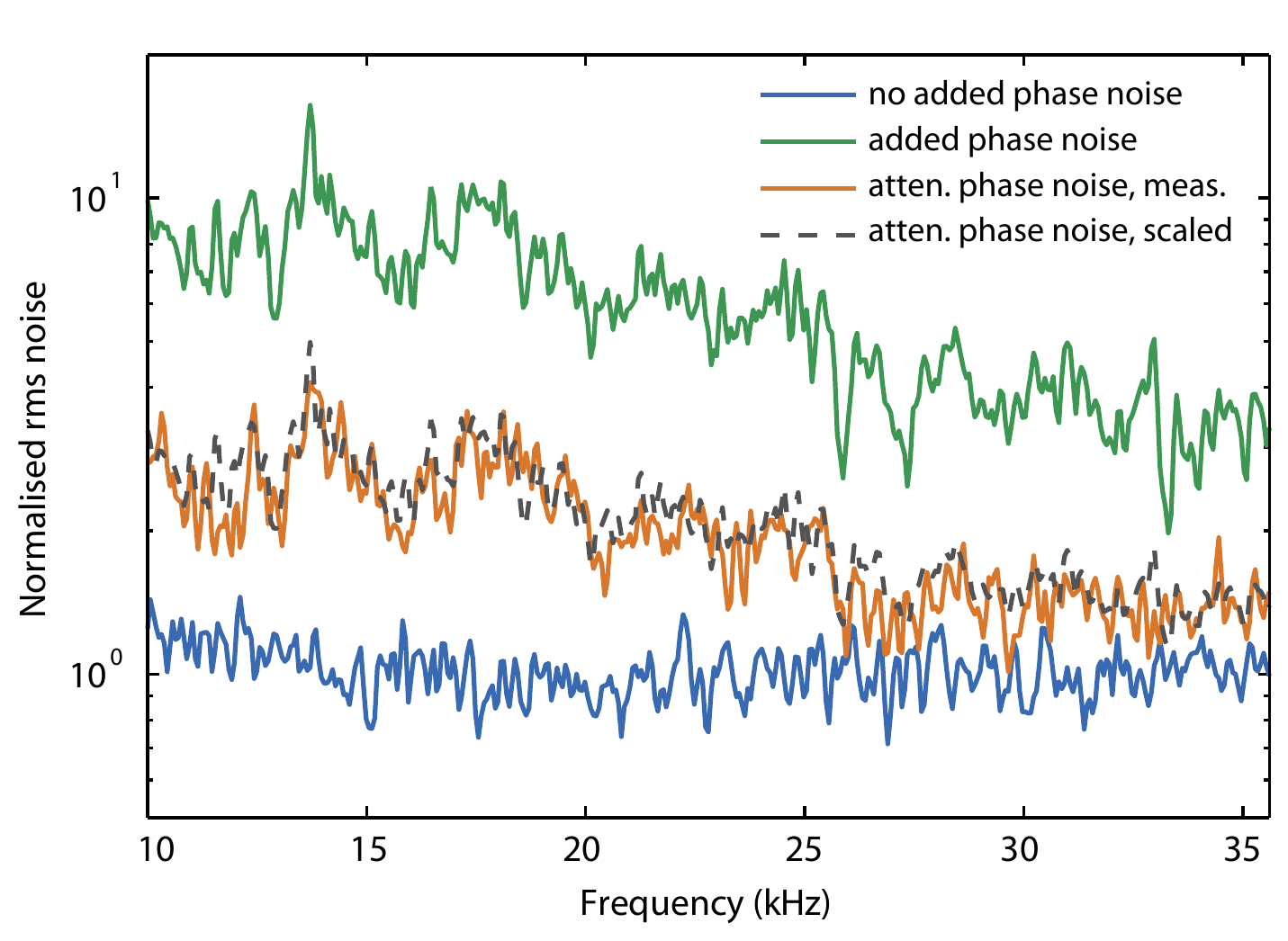}
    \caption{Experimental demonstration of the effect of phase noise on the local oscillator beam, and its dependence on signal beam power. The traces were measured with an FFT analyser as $V_\text{rms}$, averaged 10 times and then normalised to the shot-noise level.}
    \label{fig:phase_noise}
\end{figure}

To verify this, we set up a small BHD experiment in which additional phase noise was injected into the local oscillator path. The local oscillator power was set to \SI{0.55}{\milli\watt} (equivalent to $\SI{4.12}{\volt_{DC}}$ on each photodiode). We injected a small amount of laser light into the signal beam, measuring less than $\SI{3}{\micro\watt}$ or about a factor of 180 below the LO power level. The beat signal between LO and signal fields was used as an error signal to stabilise the BHD readout to the phase quadrature via a PZT-mounted mirror in the LO path. The resulting output noise spectrum was recorded with a spectrum analyser. \figref{phase_noise} shows the spectrum (in blue) obtained for no additional phase noise, normalised such that the shot-noise level is at 1. In the next step, white noise with an amplitude of $\SI{100}{\milli\volt_{pp}}$ was added into the PZT drive voltage, giving the green trace. The trace's slope can be attributed to the declining transfer function of the PZT and its high-voltage amplifier. The excess phase noise on the LO is clearly visible. After further attenuating the signal beam with a neutral density filter with transmission $T\approx 10\%$, the excess phase noise is significantly reduced (orange trace), which demonstrates that it scales with the signal power. Indeed, scaling the green curve with $\sqrt{T}$ (and taking the shot-noise contribution into account) results in the dashed grey curve, which closely fits the orange curve.

%%% =========================================================================

\section{Balanced-Homodyne Readout in Advanced LIGO}

As we have seen in Section~\ref{sec:introduction}, balanced homodyne detection is a vital prerequisite for many quantum-noise reduction techniques. In addition, balanced homodyne readout offers the possibility to reduce technical noise limitations, such as photodiode detection noise and scattered light as we will discuss in the following paragraph.

In a foreseen upgrade to the Advanced LIGO gravitational wave detectors, frequency-dependent squeezed light will be used to improve the detectors' quantum-noise limited sensitivity \cite{LSCWhitePaper2015}. Still, the readout quadrature will remain the amplitude quadrature, such that conventional DC readout seems sufficient. However, as Fritschel~\emph{et~al.}~\cite{Fritschel2014} have pointed out, it will be increasingly difficult to maintain sufficient noise margin between the photo-diode electronic noise and the shot noise for high levels of squeezing. This margin is directly related to the DC voltage level after the photo-diode transimpedance amplifier. In the presence of squeezing levels of about \SI{10}{dB}, challenging voltage levels exceeding $\SI{50}{V_{DC}}$ will be encountered. Balanced-homodyne detection would help in that the large DC offset can be directly subtracted before reaching the transimpedance amplifier stage \cite{*[{Alternative investigations are under way on how to reduce the DC current in the transimpedance stage of photo diode electronics, see }] [{}] Grote2015a}. Further technical benefits of balanced-homodyne readout are connected to the reduction of light in the output port, because the dark-fringe offset that is necessary for DC readout can be eliminated. This would help mitigating scattered light noise which originates from the output optics \cite{Martynov2015}, as well as reducing first-order coupling of beam-pointing noise on differential-wavefront sensors \cite{Grote2010}.

\begin{figure}
  \begin{minipage}[b]{0.5\linewidth}
  \includegraphics[height=7cm]{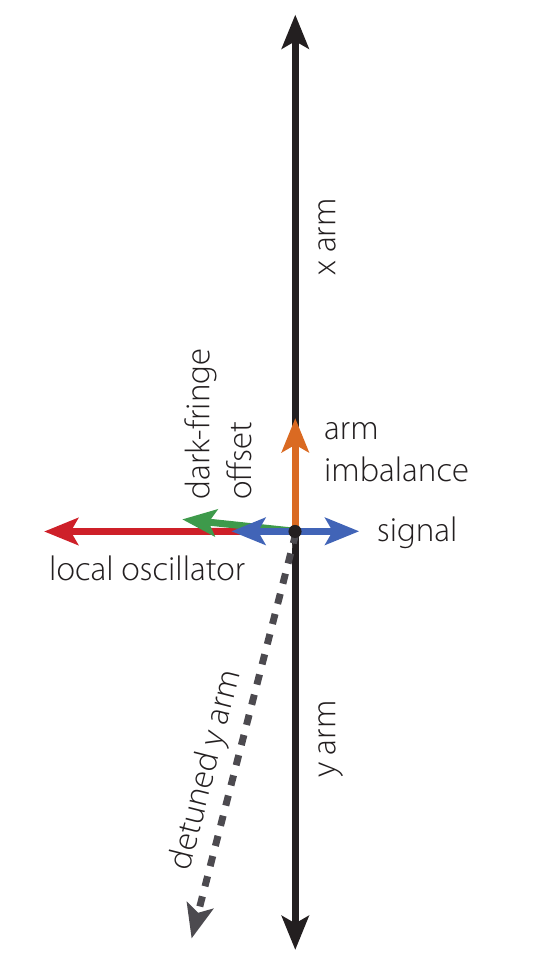}
  \end{minipage}\begin{minipage}[b]{0.5\linewidth}
    \caption{Illustration of fields in the detection port. An imbalance in the arms leads to leakage light in the output port (orange) which is orthogonal to a gravitational-wave signal (blue). A slight detuning of one of the arms leads to a dark-fringe offset field (green) in the gravitational-wave signal's quadrature. A dark-fringe offset will thus amplify amplitude fluctuations of the local oscillator (red), while an imbalance in the arms amplifies phase fluctuations.}\label{fig:phasor_diagram}
  \end{minipage}
\end{figure}

The DC signal power in Michelson-type interferometers such as Advanced LIGO is the result of two different processes, see \figref{phasor_diagram}. Firstly, the light recombining at the main beam splitter can be offset in phase by slightly detuning the length of one of the interferometer arms. This dark-fringe offset leads to light in the gravitational-wave signal's quadrature. Secondly, the recombining light can have a mismatch in amplitude, leading to some leakage light in the output field which is in the orthogonal quadrature to the gravitational wave signal. In total, this leads to a DC signal field in the output port that is oriented in some mixed quadrature orientation and both amplitude and phase quadrature noise couplings as discussed above have to be considered.

In Advanced LIGO, a local oscillator could be obtained from one of the beams that are reflected at the main beam splitter anti-reflective coating \cite{Fritschel2014}. The interferometer main laser is already amplitude stabilised to a very high degree \cite{Kwee2012}, in addition to being frequency-stabilised and filtered by the \SI{4}{km}-long power-recycling cavity. Thus, it should not be a problem to satisfy Eq.~\eqref{eq:rin_requirement} for any remaining DC signal power that can reasonably be expected to occur in the interferometer output port.

The path-length stability from Eq.~\eqref{eq:pn-req-x} sets a requirement on the suspension and control of any auxiliary optic that is not common to both local oscillator and signal path. Assuming that the signal DC power can be kept below \SI{1}{mW}, a displacement stability of about $\SI{1e-15}{\meter/\sqrthz}$ would need to be achieved. This is beyond the capabilities of simple auxiliary suspensions, but within the requirements for e.g.\ the input mode-cleaner suspensions \cite{Fritschel2015}.

%%% =========================================================================

\section{Balanced-Homodyne Readout in the Glasgow Sagnac Speed Meter}
\label{sec:glasgow_ssm}

\begin{figure}
  \includegraphics[width=\linewidth]{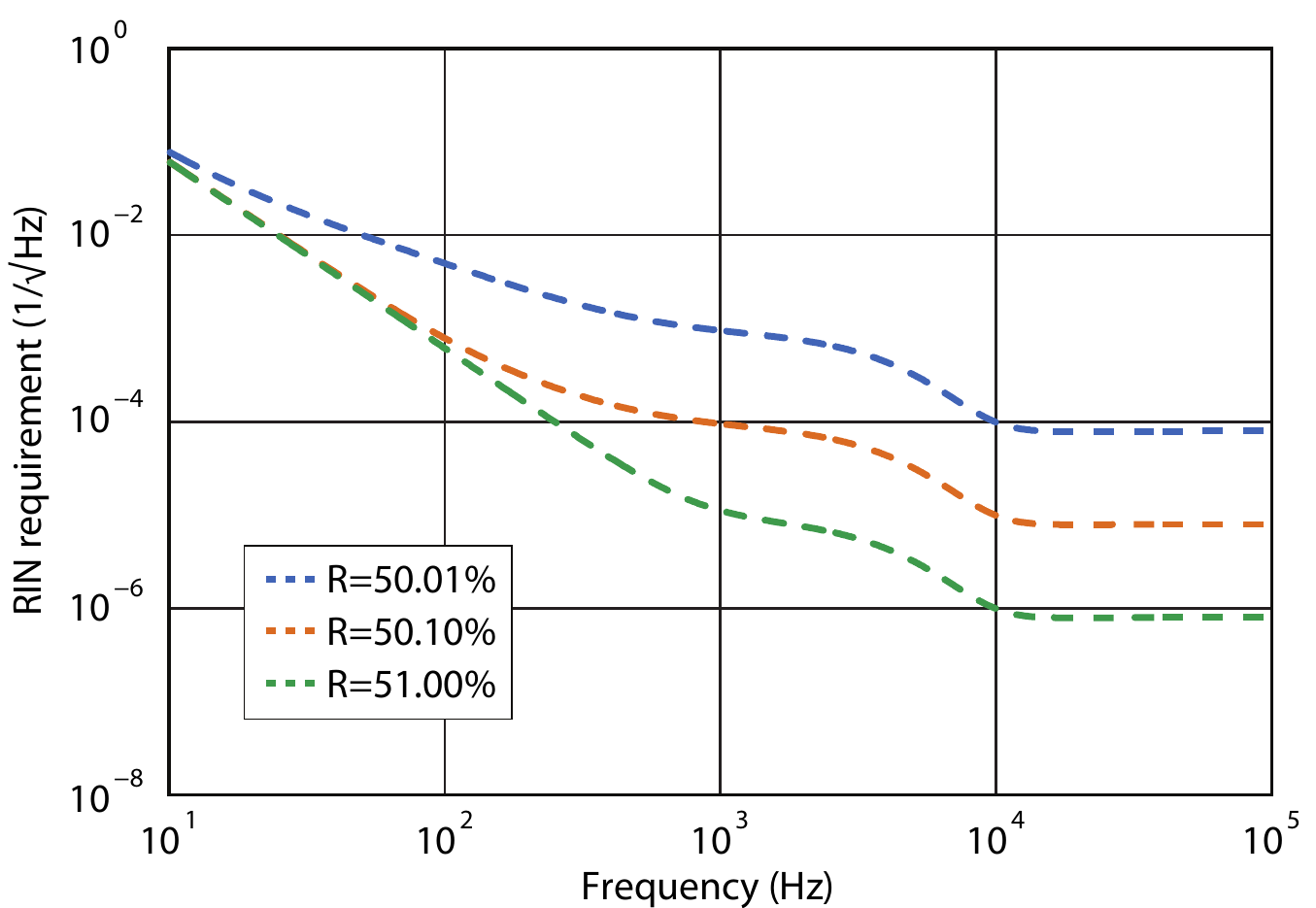}
  \caption{Simulated requirements on the local oscillator amplitude stability for the Glasgow Sagnac Speed Meter experiment, depending on main beam-splitter imbalance. The traces were simulated with Finesse \cite{Freise2004} and do not include noise sources other than quantum noise for the sensitivity modelling.}
  \label{fig:ssm_rin}
\end{figure}

The Glasgow Sagnac Speed Meter Experiment \cite{Graef2014} is a small-scale experiment with the goal of showing quantum back-action noise suppression in the Sagnac configuration \cite{Chen2003}, i.e.\ to demonstrate the feasibility of the speed meter concept for gravitational wave detectors. To achieve this goal, the experiment must be strongly dominated by radiation pressure forces at the frequencies of interest. This will be achieved by using light mirrors ($\approx \SI1\gram$), suspended from multi-stage pendulums, in combination with high laser powers (several kilowatts inside high-finesse cavities).

For an ideal Sagnac interferometer, the output port is always at a dark fringe. In addition, the output signal appears in the phase quadrature. Therefore the Sagnac interferometer is unsuitable for the DC readout scheme used in Michelson interferometers \cite{*[{There are however some clever proposals on how to produce the required local oscillator in the phase quadrature, see }] [{}] Wang2013}. Instead, the Sagnac interferometer is an obvious candidate for balanced-homodyne readout, which can be tuned to any readout quadrature and provides the necessary local oscillator.

Asymmetries in the interferometer can couple light into the output port, e.g.\ a non-perfect splitting ratio of the main beam splitter \cite{Wang2013}. The implications of such asymmetries on the Sagnac speed meter's quantum-noise limited sensitivity (QNLS) are discussed in \cite{Danilishin2015}.

In \figref{ssm_rin}, we show the impact of beam splitter imbalance on the RIN requirements for the local oscillator beam. The high-frequency level is given by the RIN of the residual DC signal power, additionally relaxed by about one order of magnitude because the Sagnac's readout is set (close) to the phase quadrature, seeing little influence of amplitude fluctuations.Towards lower frequencies, the requirements become significantly relaxed. We can therefore conclude that for the Glasgow Sagnac Speed Meter, the local oscillator does not have to be amplitude-stabilised beyond what is needed for the main laser.

In terms of path stability, we assume that a beam splitting imbalance of $\pm 0.2\%$ is achievable. This would result in about $\SI{20}{\micro\watt}$ DC signal power. According to Eq. \eqref{eq:pn-req-x} this leads to a displacement noise requirement of $< \SI{5.8e-14}{m/\sqrthz}$ in the detection band from \SIrange{100}{1000}{Hz}. We believe this is within the capabilities of our auxiliary beam steering suspensions, which we designed to achieve a horizontal displacement noise of about $\SI{1e-15}{m/\sqrthz}$ at \SI{100}{Hz}.

%%% =========================================================================

\section{Summary}

Novel aspects of quantum-noise reduction techniques in advanced gravitational wave detectors rely on a robust method for detecting arbitrary quadratures of the interferometer output field. Here we have investigated the challenges that arise when balanced homodyne readout is introduced to these detectors. In particular, we derived requirements for the amplitude noise and path length stability of the local oscillator field. We found that residual power in the signal path, caused by a contrast defect of the main interferometer, is the main driver of these requirements. We verified this result in a table-top experiment, where we introduced additional phase noise in the local oscillator path. Within the examples of Advanced LIGO and the Glasgow Sagnac Speed Meter we showed that the derived requirements are manageable with current technology and do not exclude the application of balanced homodyne readout in gravitational wave detectors.

\section{Acknowledgements}

The authors are grateful to Hartmut Grote for very valuable comments on this manuscript. The work described in this article is funded by the European Research Council (ERC-2012-StG: 307245). SS was funded by the Alexander von Humboldt Society and the International Max Planck Partnership (IMPP). We are grateful for support from Science and Technology Facilities Council (Grant Ref: ST/L000946/1) and the ASPERA ET-R\&D project.

%%% =========================================================================

\appendix

\section{Detailed calculation of balanced-homodyne detection}
\label{sec:detailed_calculations}

\newcommand{\eiphi}{\ensuremath{\re^{\ri\phi}}}
\newcommand{\emiphi}{\ensuremath{\re^{-\ri\phi}}}
Let us denote the two fields impinging onto the 50:50 beam splitter with $\op a$ and $\op b$. The relative phase between the two fields shall be $\phi$, which we write as an explicit phase $\op a \to \op a \eiphi$. The two output fields $\op c$ and $\op d$ are then given by
\begin{align}
  \op c = \frac{1}{\sqrt{2}}\left(\op a \eiphi+ \op b \right)\\
  \op d = \frac{1}{\sqrt{2}}\left(-\op a \eiphi + \op b \right)\,.
\end{align}
The light power in each output field is proportional to the number operators $\opd c \op c$ and $\opd d \op d$, respectively. These are given by
\begin{align}
  \opd c \op c &= \frac{1}{2}\left[ 
    \bigl( \opd a\emiphi + \opd b \bigr)\bigl( \op a\eiphi + \op b \bigr)
  \right] \\
    &= \frac{1}{2}\left[ \opd a \op a + \opd b \op b +
      \opd a \op b \emiphi + \op a \opd b \eiphi \right] \label{eq:cdc}
\intertext{and}
  \opd d \op d &= \frac{1}{2}\left[ 
    \bigl( -\opd a\emiphi + \opd b \bigr)\bigl( -\op a\eiphi + \op b \bigr)
  \right] \\
    &= \frac{1}{2}\left[ \opd a \op a + \opd b \op b -
      \opd a \op b \emiphi - \op a \opd b \eiphi \right]\,. \label{eq:ddd}
\end{align}
We now decompose the operators into a classical amplitude and the quantum-mechanical fluctuations, $\op a = \alpha + \delta\op{a}$. Furthermore, we linearise the equations such that terms that are quadratic in the noise vanish, $\delta\op a \delta\op b \to 0$. With this, Eq.~\eqref{eq:cdc} becomes
\begin{align}
  \opd c \op c &= \frac{1}{2}\Bigl[
      \alpha^2 + \alpha\bigl(\underbrace{\delta\op a + \delta\opd a}_{2\delta\op X_1^a} + \underbrace{\delta\op b \emiphi + \delta\opd b\eiphi}_{2\delta\op X_{-\phi}^b}\bigr) \\
    &\quad + \beta^2 + \beta\bigl(\underbrace{\delta\op b + \delta\opd b}_{2\delta\op X_1^b} + \underbrace{\delta\op a \eiphi + \delta\opd a \emiphi}_{2\delta\op X_{\phi}^a} \bigr) \\
      &\quad + \underbrace{\alpha\beta\eiphi + \alpha\beta\emiphi}_{2\alpha\beta\cos\phi} \Bigr] \\
  &= \frac{1}{2}\Bigl[\alpha^2 + \beta^2 + 2\alpha\beta\cos\phi + 2\alpha\delta\op X_1^a + 2\beta\delta\op X_1^b \\
  &\quad + 2\alpha\delta\op X_{-\phi}^b + 2\beta\delta\op X_{\phi}^a \Bigr]\,.
\end{align}
Similarly, Eq.~\eqref{eq:ddd} becomes
\begin{align}
  \opd d \op d &= \frac{1}{2}\Bigl[\alpha^2 + \beta^2 - 2\alpha\beta\cos\phi + 2\alpha\delta\op X_1^a \\
  &\quad + 2\beta\delta\op X_1^b - 2\alpha\delta\op X_{-\phi}^b - 2\beta\delta\op X_{\phi}^a \Bigr]\,.
\end{align}
The difference in photo currents is therefore proportional to
\begin{align}
  \op i_- = \opd c \op c - \opd d \op d = 2\alpha\beta\cos\phi + 2\alpha\delta\op X_{-\phi}^b + 2\beta\delta\op X_{\phi}^a\,, \label{eq:photocurrent}
\end{align}
with a noise variance of
\begin{align}
  \var\op i_- = 4\alpha^2\var\delta\op X_{-\phi}^b + 4\beta^2\var\delta\op X_{\phi}^a\,.
\end{align}

\section{Imbalanced subtraction}
\label{sec:cmrr}
We can model the effects of a non-ideal common-mode rejection ratio by an imperfect subtraction in Eq.~\eqref{eq:photocurrent}, i.e.\ by writing
\begin{align}
  \op i_{-,\epsilon} &= \opd c \op c - (1-\epsilon) \opd d \op d \\
  & = \text{const} + \epsilon\bigl( \alpha\delta\op X_1^a + \beta\delta\op X_1^b \bigr) \\
  &\quad + (2-\epsilon) \bigl( \alpha\delta\op X_{-\phi}^b + \beta\delta\op X_{\phi}^a \bigr) \,.
\end{align}
Comparing this result to Eq.\eqref{eq:photocurrent}, we see that the an additional noise term proportional to the amplitude noise $\delta\op X_1$ in the two beams is introduced. Since $\alpha \gg \beta$ and often also $\var\delta\op X_1^a \gg \var\delta\op X_1^b$, the contribution from the local oscillator $\op a$ will be dominating.

\bibliography{bhd_arxiv}

\begin{thebibliography}{32}%
\makeatletter
\providecommand \@ifxundefined [1]{%
 \@ifx{#1\undefined}
}%
\providecommand \@ifnum [1]{%
 \ifnum #1\expandafter \@firstoftwo
 \else \expandafter \@secondoftwo
 \fi
}%
\providecommand \@ifx [1]{%
 \ifx #1\expandafter \@firstoftwo
 \else \expandafter \@secondoftwo
 \fi
}%
\providecommand \natexlab [1]{#1}%
\providecommand \enquote  [1]{``#1''}%
\providecommand \bibnamefont  [1]{#1}%
\providecommand \bibfnamefont [1]{#1}%
\providecommand \citenamefont [1]{#1}%
\providecommand \href@noop [0]{\@secondoftwo}%
\providecommand \href [0]{\begingroup \@sanitize@url \@href}%
\providecommand \@href[1]{\@@startlink{#1}\@@href}%
\providecommand \@@href[1]{\endgroup#1\@@endlink}%
\providecommand \@sanitize@url [0]{\catcode `\\12\catcode `\$12\catcode
  `\&12\catcode `\#12\catcode `\^12\catcode `\_12\catcode `\%12\relax}%
\providecommand \@@startlink[1]{}%
\providecommand \@@endlink[0]{}%
\providecommand \url  [0]{\begingroup\@sanitize@url \@url }%
\providecommand \@url [1]{\endgroup\@href {#1}{\urlprefix }}%
\providecommand \urlprefix  [0]{URL }%
\providecommand \Eprint [0]{\href }%
\providecommand \doibase [0]{http://dx.doi.org/}%
\providecommand \selectlanguage [0]{\@gobble}%
\providecommand \bibinfo  [0]{\@secondoftwo}%
\providecommand \bibfield  [0]{\@secondoftwo}%
\providecommand \translation [1]{[#1]}%
\providecommand \BibitemOpen [0]{}%
\providecommand \bibitemStop [0]{}%
\providecommand \bibitemNoStop [0]{.\EOS\space}%
\providecommand \EOS [0]{\spacefactor3000\relax}%
\providecommand \BibitemShut  [1]{\csname bibitem#1\endcsname}%
\let\auto@bib@innerbib\@empty
%</preamble>
\bibitem [{\citenamefont {Einstein}(1918)}]{Einstein18}%
  \BibitemOpen
  \bibfield  {author} {\bibinfo {author} {\bibfnamefont {A.}~\bibnamefont
  {Einstein}},\ }\href@noop {} {\bibfield  {journal} {\bibinfo  {journal}
  {Sitzungsberichte der preu{\ss}ischen Akademie der Wissenschaften}\ ,\
  \bibinfo {pages} {154}} (\bibinfo {year} {1918})}\BibitemShut {NoStop}%
\bibitem [{\citenamefont {{The LIGO Scientific Collaboration}}\ \emph
  {et~al.}(2015)\citenamefont {{The LIGO Scientific Collaboration}},
  \citenamefont {Aasi}, \citenamefont {Abbott}, \citenamefont {Abbott},
  \citenamefont {Abbott}, \citenamefont {Abernathy}, \citenamefont {Ackley},
  \citenamefont {Adams}, \citenamefont {Adams}, \citenamefont {Addesso},
  \citenamefont {Adhikari}, \citenamefont {Adya}, \citenamefont {Affeldt},
  \citenamefont {Aggarwal}, \citenamefont {Aguiar} \emph
  {et~al.}}]{AdvLigo2015}%
  \BibitemOpen
  \bibfield  {author} {\bibinfo {author} {\bibnamefont {{The LIGO Scientific
  Collaboration}}}, \bibinfo {author} {\bibfnamefont {J.}~\bibnamefont {Aasi}},
  \bibinfo {author} {\bibfnamefont {B.~P.}\ \bibnamefont {Abbott}}, \bibinfo
  {author} {\bibfnamefont {R.}~\bibnamefont {Abbott}}, \bibinfo {author}
  {\bibfnamefont {T.}~\bibnamefont {Abbott}}, \bibinfo {author} {\bibfnamefont
  {M.~R.}\ \bibnamefont {Abernathy}}, \bibinfo {author} {\bibfnamefont
  {K.}~\bibnamefont {Ackley}}, \bibinfo {author} {\bibfnamefont
  {C.}~\bibnamefont {Adams}}, \bibinfo {author} {\bibfnamefont
  {T.}~\bibnamefont {Adams}}, \bibinfo {author} {\bibfnamefont
  {P.}~\bibnamefont {Addesso}}, \bibinfo {author} {\bibfnamefont {R.~X.}\
  \bibnamefont {Adhikari}}, \bibinfo {author} {\bibfnamefont {V.}~\bibnamefont
  {Adya}}, \bibinfo {author} {\bibfnamefont {C.}~\bibnamefont {Affeldt}},
  \bibinfo {author} {\bibfnamefont {N.}~\bibnamefont {Aggarwal}}, \bibinfo
  {author} {\bibfnamefont {O.~D.}\ \bibnamefont {Aguiar}},  \emph {et~al.},\
  }\href {http://stacks.iop.org/0264-9381/32/i=7/a=074001} {\bibfield
  {journal} {\bibinfo  {journal} {Classical and Quantum Gravity}\ }\textbf
  {\bibinfo {volume} {32}},\ \bibinfo {pages} {074001} (\bibinfo {year}
  {2015})}\BibitemShut {NoStop}%
\bibitem [{\citenamefont {{The LIGO Scientific Collaboration}}\ \emph
  {et~al.}(2011)\citenamefont {{The LIGO Scientific Collaboration}},
  \citenamefont {Abadie}, \citenamefont {Abbott}, \citenamefont {R.},
  \citenamefont {Abbott}, \citenamefont {Abernathy}, \citenamefont {Adams},
  \citenamefont {Adhikari}, \citenamefont {Affeldt}, \citenamefont {Allen},
  \citenamefont {Allen}, \citenamefont {Amador~Ceron}, \citenamefont
  {Amariutei}, \citenamefont {Amin} \emph {et~al.}}]{Abadie2011}%
  \BibitemOpen
  \bibfield  {author} {\bibinfo {author} {\bibnamefont {{The LIGO Scientific
  Collaboration}}}, \bibinfo {author} {\bibfnamefont {J.}~\bibnamefont
  {Abadie}}, \bibinfo {author} {\bibfnamefont {B.~P.}\ \bibnamefont {Abbott}},
  \bibinfo {author} {\bibfnamefont {A.}~\bibnamefont {R.}}, \bibinfo {author}
  {\bibfnamefont {T.~D.}\ \bibnamefont {Abbott}}, \bibinfo {author}
  {\bibfnamefont {M.}~\bibnamefont {Abernathy}}, \bibinfo {author}
  {\bibfnamefont {C.}~\bibnamefont {Adams}}, \bibinfo {author} {\bibfnamefont
  {R.}~\bibnamefont {Adhikari}}, \bibinfo {author} {\bibfnamefont
  {C.}~\bibnamefont {Affeldt}}, \bibinfo {author} {\bibfnamefont
  {B.}~\bibnamefont {Allen}}, \bibinfo {author} {\bibfnamefont {G.~S.}\
  \bibnamefont {Allen}}, \bibinfo {author} {\bibfnamefont {E.}~\bibnamefont
  {Amador~Ceron}}, \bibinfo {author} {\bibfnamefont {D.}~\bibnamefont
  {Amariutei}}, \bibinfo {author} {\bibfnamefont {R.~S.}\ \bibnamefont {Amin}},
   \emph {et~al.},\ }\href {\doibase 10.1038/nphys2083} {\bibfield  {journal}
  {\bibinfo  {journal} {Nature Physics}\ }\textbf {\bibinfo {volume} {7}},\
  \bibinfo {pages} {962} (\bibinfo {year} {2011})}\BibitemShut {NoStop}%
\bibitem [{\citenamefont {{The LIGO Scientific Collaboration}}\ \emph
  {et~al.}(2013)\citenamefont {{The LIGO Scientific Collaboration}},
  \citenamefont {Aasi}, \citenamefont {Abadie}, \citenamefont {Abbott},
  \citenamefont {Abbott}, \citenamefont {Abbott}, \citenamefont {Abernathy},
  \citenamefont {Adams}, \citenamefont {Adams}, \citenamefont {Addesso},
  \citenamefont {Adhikari} \emph {et~al.}}]{Aasi2013}%
  \BibitemOpen
  \bibfield  {author} {\bibinfo {author} {\bibnamefont {{The LIGO Scientific
  Collaboration}}}, \bibinfo {author} {\bibfnamefont {J.}~\bibnamefont {Aasi}},
  \bibinfo {author} {\bibfnamefont {J.}~\bibnamefont {Abadie}}, \bibinfo
  {author} {\bibfnamefont {B.}~\bibnamefont {Abbott}}, \bibinfo {author}
  {\bibfnamefont {R.}~\bibnamefont {Abbott}}, \bibinfo {author} {\bibfnamefont
  {T.}~\bibnamefont {Abbott}}, \bibinfo {author} {\bibfnamefont
  {M.}~\bibnamefont {Abernathy}}, \bibinfo {author} {\bibfnamefont
  {C.}~\bibnamefont {Adams}}, \bibinfo {author} {\bibfnamefont
  {T.}~\bibnamefont {Adams}}, \bibinfo {author} {\bibfnamefont
  {P.}~\bibnamefont {Addesso}}, \bibinfo {author} {\bibfnamefont
  {R.}~\bibnamefont {Adhikari}},  \emph {et~al.},\ }\href@noop {} {\bibfield
  {journal} {\bibinfo  {journal} {Nature Photonics}\ }\textbf {\bibinfo
  {volume} {7}},\ \bibinfo {pages} {613} (\bibinfo {year} {2013})}\BibitemShut
  {NoStop}%
\bibitem [{\citenamefont {Grote}\ \emph {et~al.}(2013)\citenamefont {Grote},
  \citenamefont {Danzmann}, \citenamefont {Dooley}, \citenamefont {Schnabel},
  \citenamefont {Slutsky},\ and\ \citenamefont {Vahlbruch}}]{Grote2013}%
  \BibitemOpen
  \bibfield  {author} {\bibinfo {author} {\bibfnamefont {H.}~\bibnamefont
  {Grote}}, \bibinfo {author} {\bibfnamefont {K.}~\bibnamefont {Danzmann}},
  \bibinfo {author} {\bibfnamefont {K.~L.}\ \bibnamefont {Dooley}}, \bibinfo
  {author} {\bibfnamefont {R.}~\bibnamefont {Schnabel}}, \bibinfo {author}
  {\bibfnamefont {J.}~\bibnamefont {Slutsky}}, \ and\ \bibinfo {author}
  {\bibfnamefont {H.}~\bibnamefont {Vahlbruch}},\ }\href {\doibase
  10.1103/PhysRevLett.110.181101} {\bibfield  {journal} {\bibinfo  {journal}
  {Phys. Rev. Lett.}\ }\textbf {\bibinfo {volume} {110}},\ \bibinfo {pages}
  {181101} (\bibinfo {year} {2013})}\BibitemShut {NoStop}%
\bibitem [{\citenamefont {Kimble}\ \emph {et~al.}(2001)\citenamefont {Kimble},
  \citenamefont {Levin}, \citenamefont {Matsko}, \citenamefont {Thorne},\ and\
  \citenamefont {Vyatchanin}}]{Kimble01}%
  \BibitemOpen
  \bibfield  {author} {\bibinfo {author} {\bibfnamefont {H.~J.}\ \bibnamefont
  {Kimble}}, \bibinfo {author} {\bibfnamefont {Y.}~\bibnamefont {Levin}},
  \bibinfo {author} {\bibfnamefont {A.~B.}\ \bibnamefont {Matsko}}, \bibinfo
  {author} {\bibfnamefont {K.~S.}\ \bibnamefont {Thorne}}, \ and\ \bibinfo
  {author} {\bibfnamefont {S.~P.}\ \bibnamefont {Vyatchanin}},\ }\href
  {\doibase 10.1103/PhysRevD.65.022002} {\bibfield  {journal} {\bibinfo
  {journal} {Physical Review D}\ }\textbf {\bibinfo {volume} {65}},\ \bibinfo
  {pages} {22002} (\bibinfo {year} {2001})}\BibitemShut {NoStop}%
\bibitem [{\citenamefont {Braginsky}\ \emph {et~al.}(1980)\citenamefont
  {Braginsky}, \citenamefont {Vorontsov},\ and\ \citenamefont
  {Thorne}}]{Braginsky1980}%
  \BibitemOpen
  \bibfield  {author} {\bibinfo {author} {\bibfnamefont {V.~B.}\ \bibnamefont
  {Braginsky}}, \bibinfo {author} {\bibfnamefont {Y.~I.}\ \bibnamefont
  {Vorontsov}}, \ and\ \bibinfo {author} {\bibfnamefont {K.~S.}\ \bibnamefont
  {Thorne}},\ }\href {\doibase 10.1126/science.209.4456.547} {\bibfield
  {journal} {\bibinfo  {journal} {Science}\ }\textbf {\bibinfo {volume}
  {209}},\ \bibinfo {pages} {547} (\bibinfo {year} {1980})}\BibitemShut
  {NoStop}%
\bibitem [{\citenamefont {{Braginsky}}\ and\ \citenamefont
  {{Khalili}}(1990)}]{Braginsky90}%
  \BibitemOpen
  \bibfield  {author} {\bibinfo {author} {\bibfnamefont {V.~B.}\ \bibnamefont
  {{Braginsky}}}\ and\ \bibinfo {author} {\bibfnamefont {F.~J.}\ \bibnamefont
  {{Khalili}}},\ }\href {\doibase 10.1016/0375-9601(90)90442-Q} {\bibfield
  {journal} {\bibinfo  {journal} {Physics Letters A}\ }\textbf {\bibinfo
  {volume} {147}},\ \bibinfo {pages} {251} (\bibinfo {year}
  {1990})}\BibitemShut {NoStop}%
\bibitem [{\citenamefont {Chen}(2003)}]{Chen2003}%
  \BibitemOpen
  \bibfield  {author} {\bibinfo {author} {\bibfnamefont {Y.}~\bibnamefont
  {Chen}},\ }\href {\doibase 10.1103/PhysRevD.67.122004} {\bibfield  {journal}
  {\bibinfo  {journal} {Phys. Rev. D}\ }\textbf {\bibinfo {volume} {67}},\
  \bibinfo {pages} {122004} (\bibinfo {year} {2003})}\BibitemShut {NoStop}%
\bibitem [{\citenamefont {Gr\"af}\ \emph {et~al.}(2014)\citenamefont {Gr\"af},
  \citenamefont {Barr}, \citenamefont {Bell}, \citenamefont {Campbell},
  \citenamefont {Cumming}, \citenamefont {Danilishin}, \citenamefont {Gordon},
  \citenamefont {Hammond}, \citenamefont {Hennig}, \citenamefont {Houston},
  \citenamefont {Huttner}, \citenamefont {Jones}, \citenamefont {Leavey},
  \citenamefont {L{\"u}ck}, \citenamefont {Macarthur}, \citenamefont {Marwick},
  \citenamefont {Rigby}, \citenamefont {Schilling}, \citenamefont {Sorazu},
  \citenamefont {Spencer}, \citenamefont {Steinlechner}, \citenamefont
  {Strain},\ and\ \citenamefont {Hild}}]{Graef2014}%
  \BibitemOpen
  \bibfield  {author} {\bibinfo {author} {\bibfnamefont {C.}~\bibnamefont
  {Gr\"af}}, \bibinfo {author} {\bibfnamefont {B.~W.}\ \bibnamefont {Barr}},
  \bibinfo {author} {\bibfnamefont {A.~S.}\ \bibnamefont {Bell}}, \bibinfo
  {author} {\bibfnamefont {F.}~\bibnamefont {Campbell}}, \bibinfo {author}
  {\bibfnamefont {A.~V.}\ \bibnamefont {Cumming}}, \bibinfo {author}
  {\bibfnamefont {S.~L.}\ \bibnamefont {Danilishin}}, \bibinfo {author}
  {\bibfnamefont {N.~A.}\ \bibnamefont {Gordon}}, \bibinfo {author}
  {\bibfnamefont {G.~D.}\ \bibnamefont {Hammond}}, \bibinfo {author}
  {\bibfnamefont {J.}~\bibnamefont {Hennig}}, \bibinfo {author} {\bibfnamefont
  {E.~A.}\ \bibnamefont {Houston}}, \bibinfo {author} {\bibfnamefont {S.~H.}\
  \bibnamefont {Huttner}}, \bibinfo {author} {\bibfnamefont {R.~A.}\
  \bibnamefont {Jones}}, \bibinfo {author} {\bibfnamefont {S.~S.}\ \bibnamefont
  {Leavey}}, \bibinfo {author} {\bibfnamefont {H.}~\bibnamefont {L{\"u}ck}},
  \bibinfo {author} {\bibfnamefont {J.}~\bibnamefont {Macarthur}}, \bibinfo
  {author} {\bibfnamefont {M.}~\bibnamefont {Marwick}}, \bibinfo {author}
  {\bibfnamefont {S.}~\bibnamefont {Rigby}}, \bibinfo {author} {\bibfnamefont
  {R.}~\bibnamefont {Schilling}}, \bibinfo {author} {\bibfnamefont
  {B.}~\bibnamefont {Sorazu}}, \bibinfo {author} {\bibfnamefont
  {A.}~\bibnamefont {Spencer}}, \bibinfo {author} {\bibfnamefont
  {S.}~\bibnamefont {Steinlechner}}, \bibinfo {author} {\bibfnamefont {K.~A.}\
  \bibnamefont {Strain}}, \ and\ \bibinfo {author} {\bibfnamefont
  {S.}~\bibnamefont {Hild}},\ }\href
  {http://stacks.iop.org/0264-9381/31/i=21/a=215009} {\bibfield  {journal}
  {\bibinfo  {journal} {Classical and Quantum Gravity}\ }\textbf {\bibinfo
  {volume} {31}},\ \bibinfo {pages} {215009} (\bibinfo {year}
  {2014})}\BibitemShut {NoStop}%
\bibitem [{\citenamefont {Hild}\ \emph {et~al.}(2009)\citenamefont {Hild},
  \citenamefont {Grote}, \citenamefont {Degallaix}, \citenamefont {Chelkowski},
  \citenamefont {Danzmann}, \citenamefont {Freise}, \citenamefont {Hewitson},
  \citenamefont {Hough}, \citenamefont {L{\"u}ck}, \citenamefont {Prijatelj}
  \emph {et~al.}}]{Hild2009}%
  \BibitemOpen
  \bibfield  {author} {\bibinfo {author} {\bibfnamefont {S.}~\bibnamefont
  {Hild}}, \bibinfo {author} {\bibfnamefont {H.}~\bibnamefont {Grote}},
  \bibinfo {author} {\bibfnamefont {J.}~\bibnamefont {Degallaix}}, \bibinfo
  {author} {\bibfnamefont {S.}~\bibnamefont {Chelkowski}}, \bibinfo {author}
  {\bibfnamefont {K.}~\bibnamefont {Danzmann}}, \bibinfo {author}
  {\bibfnamefont {A.}~\bibnamefont {Freise}}, \bibinfo {author} {\bibfnamefont
  {M.}~\bibnamefont {Hewitson}}, \bibinfo {author} {\bibfnamefont
  {J.}~\bibnamefont {Hough}}, \bibinfo {author} {\bibfnamefont
  {H.}~\bibnamefont {L{\"u}ck}}, \bibinfo {author} {\bibfnamefont
  {M.}~\bibnamefont {Prijatelj}},  \emph {et~al.},\ }\href@noop {} {\bibfield
  {journal} {\bibinfo  {journal} {Classical and Quantum Gravity}\ }\textbf
  {\bibinfo {volume} {26}},\ \bibinfo {pages} {055012} (\bibinfo {year}
  {2009})}\BibitemShut {NoStop}%
\bibitem [{\citenamefont {Fricke}\ \emph {et~al.}(2012)\citenamefont {Fricke},
  \citenamefont {Smith-Lefebvre}, \citenamefont {Abbott}, \citenamefont
  {Adhikari}, \citenamefont {Dooley}, \citenamefont {Evans}, \citenamefont
  {Fritschel}, \citenamefont {Frolov}, \citenamefont {Kawabe}, \citenamefont
  {Kissel} \emph {et~al.}}]{Fricke2012}%
  \BibitemOpen
  \bibfield  {author} {\bibinfo {author} {\bibfnamefont {T.~T.}\ \bibnamefont
  {Fricke}}, \bibinfo {author} {\bibfnamefont {N.~D.}\ \bibnamefont
  {Smith-Lefebvre}}, \bibinfo {author} {\bibfnamefont {R.}~\bibnamefont
  {Abbott}}, \bibinfo {author} {\bibfnamefont {R.}~\bibnamefont {Adhikari}},
  \bibinfo {author} {\bibfnamefont {K.~L.}\ \bibnamefont {Dooley}}, \bibinfo
  {author} {\bibfnamefont {M.}~\bibnamefont {Evans}}, \bibinfo {author}
  {\bibfnamefont {P.}~\bibnamefont {Fritschel}}, \bibinfo {author}
  {\bibfnamefont {V.~V.}\ \bibnamefont {Frolov}}, \bibinfo {author}
  {\bibfnamefont {K.}~\bibnamefont {Kawabe}}, \bibinfo {author} {\bibfnamefont
  {J.~S.}\ \bibnamefont {Kissel}},  \emph {et~al.},\ }\href@noop {} {\bibfield
  {journal} {\bibinfo  {journal} {Classical and Quantum Gravity}\ }\textbf
  {\bibinfo {volume} {29}},\ \bibinfo {pages} {065005} (\bibinfo {year}
  {2012})}\BibitemShut {NoStop}%
\bibitem [{\citenamefont {Leonhardt}(1997)}]{Leonhardt1997}%
  \BibitemOpen
  \bibfield  {author} {\bibinfo {author} {\bibfnamefont {U.}~\bibnamefont
  {Leonhardt}},\ }\href@noop {} {\emph {\bibinfo {title} {Measuring the quantum
  state of light}}}\ (\bibinfo  {publisher} {Cambridge University Press},\
  \bibinfo {year} {1997})\BibitemShut {NoStop}%
\bibitem [{\citenamefont {DiGuglielmo}\ \emph {et~al.}(2011)\citenamefont
  {DiGuglielmo}, \citenamefont {Samblowski}, \citenamefont {Hage},
  \citenamefont {Pineda}, \citenamefont {Eisert},\ and\ \citenamefont
  {Schnabel}}]{DiGuglielmo2011}%
  \BibitemOpen
  \bibfield  {author} {\bibinfo {author} {\bibfnamefont {J.}~\bibnamefont
  {DiGuglielmo}}, \bibinfo {author} {\bibfnamefont {A.}~\bibnamefont
  {Samblowski}}, \bibinfo {author} {\bibfnamefont {B.}~\bibnamefont {Hage}},
  \bibinfo {author} {\bibfnamefont {C.}~\bibnamefont {Pineda}}, \bibinfo
  {author} {\bibfnamefont {J.}~\bibnamefont {Eisert}}, \ and\ \bibinfo {author}
  {\bibfnamefont {R.}~\bibnamefont {Schnabel}},\ }\href {\doibase
  10.1103/PhysRevLett.107.240503} {\bibfield  {journal} {\bibinfo  {journal}
  {Physical Review Letters}\ }\textbf {\bibinfo {volume} {107}},\ \bibinfo
  {pages} {1} (\bibinfo {year} {2011})}\BibitemShut {NoStop}%
\bibitem [{\citenamefont {Bachor}\ and\ \citenamefont
  {Ralph}(2004)}]{Bachor2004}%
  \BibitemOpen
  \bibfield  {author} {\bibinfo {author} {\bibfnamefont {H.-A.}\ \bibnamefont
  {Bachor}}\ and\ \bibinfo {author} {\bibfnamefont {T.~C.}\ \bibnamefont
  {Ralph}},\ }\href@noop {} {\emph {\bibinfo {title} {A guide to experiments in
  Quantum Optics, 2nd edition}}}\ (\bibinfo  {publisher} {Wiley-VCH},\ \bibinfo
  {year} {2004})\BibitemShut {NoStop}%
\bibitem [{\citenamefont {Ast}\ \emph {et~al.}(2013)\citenamefont {Ast},
  \citenamefont {Mehmet},\ and\ \citenamefont {Schnabel}}]{Ast2013}%
  \BibitemOpen
  \bibfield  {author} {\bibinfo {author} {\bibfnamefont {S.}~\bibnamefont
  {Ast}}, \bibinfo {author} {\bibfnamefont {M.}~\bibnamefont {Mehmet}}, \ and\
  \bibinfo {author} {\bibfnamefont {R.}~\bibnamefont {Schnabel}},\ }\href@noop
  {} {\bibfield  {journal} {\bibinfo  {journal} {Optics express}\ }\textbf
  {\bibinfo {volume} {21}},\ \bibinfo {pages} {13572} (\bibinfo {year}
  {2013})}\BibitemShut {NoStop}%
\bibitem [{\citenamefont {Stefszky}\ \emph {et~al.}(2012)\citenamefont
  {Stefszky}, \citenamefont {Mow-Lowry}, \citenamefont {Chua}, \citenamefont
  {Shaddock}, \citenamefont {Buchler}, \citenamefont {Vahlbruch}, \citenamefont
  {Khalaidovski}, \citenamefont {Schnabel}, \citenamefont {Lam},\ and\
  \citenamefont {McClelland}}]{Stefszky2012}%
  \BibitemOpen
  \bibfield  {author} {\bibinfo {author} {\bibfnamefont {M.}~\bibnamefont
  {Stefszky}}, \bibinfo {author} {\bibfnamefont {C.}~\bibnamefont {Mow-Lowry}},
  \bibinfo {author} {\bibfnamefont {S.}~\bibnamefont {Chua}}, \bibinfo {author}
  {\bibfnamefont {D.}~\bibnamefont {Shaddock}}, \bibinfo {author}
  {\bibfnamefont {B.}~\bibnamefont {Buchler}}, \bibinfo {author} {\bibfnamefont
  {H.}~\bibnamefont {Vahlbruch}}, \bibinfo {author} {\bibfnamefont
  {A.}~\bibnamefont {Khalaidovski}}, \bibinfo {author} {\bibfnamefont
  {R.}~\bibnamefont {Schnabel}}, \bibinfo {author} {\bibfnamefont
  {P.}~\bibnamefont {Lam}}, \ and\ \bibinfo {author} {\bibfnamefont
  {D.}~\bibnamefont {McClelland}},\ }\href@noop {} {\bibfield  {journal}
  {\bibinfo  {journal} {Classical and Quantum Gravity}\ }\textbf {\bibinfo
  {volume} {29}},\ \bibinfo {pages} {145015} (\bibinfo {year}
  {2012})}\BibitemShut {NoStop}%
\bibitem [{\citenamefont {Fritschel}\ \emph {et~al.}(2014)\citenamefont
  {Fritschel}, \citenamefont {Evans},\ and\ \citenamefont
  {Frolov}}]{Fritschel2014}%
  \BibitemOpen
  \bibfield  {author} {\bibinfo {author} {\bibfnamefont {P.}~\bibnamefont
  {Fritschel}}, \bibinfo {author} {\bibfnamefont {M.}~\bibnamefont {Evans}}, \
  and\ \bibinfo {author} {\bibfnamefont {V.}~\bibnamefont {Frolov}},\ }\href
  {http://www.opticsinfobase.org/abstract.cfm?uri=oe-22-4-4224} {\bibfield
  {journal} {\bibinfo  {journal} {Optics express}\ }\textbf {\bibinfo {volume}
  {22}},\ \bibinfo {pages} {4224} (\bibinfo {year} {2014})}\BibitemShut
  {NoStop}%
\bibitem [{\citenamefont {Gerry}\ and\ \citenamefont {Knight}(2005)}]{Gerry05}%
  \BibitemOpen
  \bibfield  {author} {\bibinfo {author} {\bibfnamefont {C.}~\bibnamefont
  {Gerry}}\ and\ \bibinfo {author} {\bibfnamefont {P.}~\bibnamefont {Knight}},\
  }\href@noop {} {\emph {\bibinfo {title} {Introductory Quantum Optics}}}\
  (\bibinfo  {publisher} {Cambridge University Press},\ \bibinfo {year}
  {2005})\BibitemShut {NoStop}%
\bibitem [{\citenamefont {Kwee}\ \emph {et~al.}(2009)\citenamefont {Kwee},
  \citenamefont {Willke},\ and\ \citenamefont {Danzmann}}]{Kwee2009}%
  \BibitemOpen
  \bibfield  {author} {\bibinfo {author} {\bibfnamefont {P.}~\bibnamefont
  {Kwee}}, \bibinfo {author} {\bibfnamefont {B.}~\bibnamefont {Willke}}, \ and\
  \bibinfo {author} {\bibfnamefont {K.}~\bibnamefont {Danzmann}},\ }\href
  {\doibase 10.1364/OL.34.002912} {\bibfield  {journal} {\bibinfo  {journal}
  {Opt. Lett.}\ }\textbf {\bibinfo {volume} {34}},\ \bibinfo {pages} {2912}
  (\bibinfo {year} {2009})}\BibitemShut {NoStop}%
\bibitem [{\citenamefont {McKenzie}\ \emph {et~al.}(2007)\citenamefont
  {McKenzie}, \citenamefont {Gray}, \citenamefont {Lam},\ and\ \citenamefont
  {McClelland}}]{McKenzie2007}%
  \BibitemOpen
  \bibfield  {author} {\bibinfo {author} {\bibfnamefont {K.}~\bibnamefont
  {McKenzie}}, \bibinfo {author} {\bibfnamefont {M.~B.}\ \bibnamefont {Gray}},
  \bibinfo {author} {\bibfnamefont {P.~K.}\ \bibnamefont {Lam}}, \ and\
  \bibinfo {author} {\bibfnamefont {D.~E.}\ \bibnamefont {McClelland}},\ }\href
  {\doibase 10.1364/AO.46.003389} {\bibfield  {journal} {\bibinfo  {journal}
  {Appl. Opt.}\ }\textbf {\bibinfo {volume} {46}},\ \bibinfo {pages} {3389}
  (\bibinfo {year} {2007})}\BibitemShut {NoStop}%
\bibitem [{\citenamefont {Freise}\ \emph {et~al.}(2004)\citenamefont {Freise},
  \citenamefont {Heinzel}, \citenamefont {L\"{u}ck}, \citenamefont {Schilling},
  \citenamefont {Willke},\ and\ \citenamefont {Danzmann}}]{Freise2004}%
  \BibitemOpen
  \bibfield  {author} {\bibinfo {author} {\bibfnamefont {A.}~\bibnamefont
  {Freise}}, \bibinfo {author} {\bibfnamefont {G.}~\bibnamefont {Heinzel}},
  \bibinfo {author} {\bibfnamefont {H.}~\bibnamefont {L\"{u}ck}}, \bibinfo
  {author} {\bibfnamefont {R.}~\bibnamefont {Schilling}}, \bibinfo {author}
  {\bibfnamefont {B.}~\bibnamefont {Willke}}, \ and\ \bibinfo {author}
  {\bibfnamefont {K.}~\bibnamefont {Danzmann}},\ }\href
  {http://stacks.iop.org/0264-9381/21/i=5/a=102} {\bibfield  {journal}
  {\bibinfo  {journal} {Classical and Quantum Gravity}\ }\textbf {\bibinfo
  {volume} {21}},\ \bibinfo {pages} {S1067} (\bibinfo {year}
  {2004})}\BibitemShut {NoStop}%
\bibitem [{\citenamefont {Grote}(2015{\natexlab{a}})}]{Grote2015}%
  \BibitemOpen
  \bibfield  {author} {\bibinfo {author} {\bibfnamefont {H.}~\bibnamefont
  {Grote}},\ }\href@noop {} {}\bibinfo {howpublished} {personal communication}
  (\bibinfo {year} {2015}{\natexlab{a}})\BibitemShut {NoStop}%
\bibitem [{\citenamefont {Danilishin}\ and\ \citenamefont
  {Khalili}(2012)}]{Danilishin2012}%
  \BibitemOpen
  \bibfield  {author} {\bibinfo {author} {\bibfnamefont {S.~L.}\ \bibnamefont
  {Danilishin}}\ and\ \bibinfo {author} {\bibfnamefont {F.~Y.}\ \bibnamefont
  {Khalili}},\ }\href {\doibase 10.12942/lrr-2012-5} {\bibfield  {journal}
  {\bibinfo  {journal} {Living Reviews in Relativity}\ }\textbf {\bibinfo
  {volume} {15}},\ \bibinfo {pages} {5} (\bibinfo {year} {2012})}\BibitemShut
  {NoStop}%
\bibitem [{\citenamefont {{LIGO Scientific
  Collaboration}}(2015)}]{LSCWhitePaper2015}%
  \BibitemOpen
  \bibfield  {author} {\bibinfo {author} {\bibnamefont {{LIGO Scientific
  Collaboration}}},\ }\href {https://dcc.ligo.org/LIGO-T1400316/public}
  {\enquote {\bibinfo {title} {Instrument science white paper},}\ } (\bibinfo
  {year} {2015}),\ \bibinfo {note} {{LIGO-T1400316}}\BibitemShut {NoStop}%
\bibitem [{\citenamefont {Grote}(2015{\natexlab{b}})}]{Grote2015a}%
  \BibitemOpen
  \bibfield  {author} {\bibinfo {author} {\bibfnamefont {H.}~\bibnamefont
  {Grote}},\ }\href {https://dcc.ligo.org/LIGO-G1500371} {\enquote {\bibinfo
  {title} {Low noise electronics for {DC} readout},}\ } (\bibinfo {year}
  {2015}{\natexlab{b}}),\ \bibinfo {note} {{LVC internal document,
  LIGO-G1500371}}\BibitemShut {NoStop}%
\bibitem [{\citenamefont {Martynov}(2015)}]{Martynov2015}%
  \BibitemOpen
  \bibfield  {author} {\bibinfo {author} {\bibfnamefont {D.}~\bibnamefont
  {Martynov}},\ }\href {https://dcc.ligo.org/LIGO-G1500656} {\enquote {\bibinfo
  {title} {Investigations of scattered light noise in {aLIGO}
  interferometers},}\ } (\bibinfo {year} {2015}),\ \bibinfo {note} {{LVC
  internal document, LIGO-G1500656}}\BibitemShut {NoStop}%
\bibitem [{\citenamefont {Grote}\ and\ \citenamefont {the LIGO
  Scientific~Collaboration}(2010)}]{Grote2010}%
  \BibitemOpen
  \bibfield  {author} {\bibinfo {author} {\bibfnamefont {H.}~\bibnamefont
  {Grote}}\ and\ \bibinfo {author} {\bibnamefont {the LIGO
  Scientific~Collaboration}},\ }\href
  {http://stacks.iop.org/0264-9381/27/i=8/a=084003} {\bibfield  {journal}
  {\bibinfo  {journal} {Classical and Quantum Gravity}\ }\textbf {\bibinfo
  {volume} {27}},\ \bibinfo {pages} {084003} (\bibinfo {year}
  {2010})}\BibitemShut {NoStop}%
\bibitem [{\citenamefont {Kwee}\ \emph {et~al.}(2012)\citenamefont {Kwee},
  \citenamefont {Bogan}, \citenamefont {Danzmann}, \citenamefont {Frede},
  \citenamefont {Kim}, \citenamefont {King}, \citenamefont {P\"{o}ld},
  \citenamefont {Puncken}, \citenamefont {Savage}, \citenamefont {Seifert},
  \citenamefont {Wessels}, \citenamefont {Winkelmann},\ and\ \citenamefont
  {Willke}}]{Kwee2012}%
  \BibitemOpen
  \bibfield  {author} {\bibinfo {author} {\bibfnamefont {P.}~\bibnamefont
  {Kwee}}, \bibinfo {author} {\bibfnamefont {C.}~\bibnamefont {Bogan}},
  \bibinfo {author} {\bibfnamefont {K.}~\bibnamefont {Danzmann}}, \bibinfo
  {author} {\bibfnamefont {M.}~\bibnamefont {Frede}}, \bibinfo {author}
  {\bibfnamefont {H.}~\bibnamefont {Kim}}, \bibinfo {author} {\bibfnamefont
  {P.}~\bibnamefont {King}}, \bibinfo {author} {\bibfnamefont {J.}~\bibnamefont
  {P\"{o}ld}}, \bibinfo {author} {\bibfnamefont {O.}~\bibnamefont {Puncken}},
  \bibinfo {author} {\bibfnamefont {R.~L.}\ \bibnamefont {Savage}}, \bibinfo
  {author} {\bibfnamefont {F.}~\bibnamefont {Seifert}}, \bibinfo {author}
  {\bibfnamefont {P.}~\bibnamefont {Wessels}}, \bibinfo {author} {\bibfnamefont
  {L.}~\bibnamefont {Winkelmann}}, \ and\ \bibinfo {author} {\bibfnamefont
  {B.}~\bibnamefont {Willke}},\ }\href {\doibase 10.1364/OE.20.010617}
  {\bibfield  {journal} {\bibinfo  {journal} {Opt. Express}\ }\textbf {\bibinfo
  {volume} {20}},\ \bibinfo {pages} {10617} (\bibinfo {year}
  {2012})}\BibitemShut {NoStop}%
\bibitem [{\citenamefont {Fritschel}\ and\ \citenamefont
  {Coyne}(2015)}]{Fritschel2015}%
  \BibitemOpen
  \bibfield  {author} {\bibinfo {author} {\bibfnamefont {P.}~\bibnamefont
  {Fritschel}}\ and\ \bibinfo {author} {\bibfnamefont {D.}~\bibnamefont
  {Coyne}},\ }\href {https://dcc.ligo.org/T010075/public} {\enquote {\bibinfo
  {title} {Advanced {LIGO} systems design},}\ } (\bibinfo {year} {2015}),\
  \bibinfo {note} {{LIGO-T010075-v3}}\BibitemShut {NoStop}%
\bibitem [{\citenamefont {Wang}\ \emph {et~al.}(2013)\citenamefont {Wang},
  \citenamefont {Bond}, \citenamefont {Brown}, \citenamefont {Br\"uckner},
  \citenamefont {Carbone}, \citenamefont {Palmer},\ and\ \citenamefont
  {Freise}}]{Wang2013}%
  \BibitemOpen
  \bibfield  {author} {\bibinfo {author} {\bibfnamefont {M.}~\bibnamefont
  {Wang}}, \bibinfo {author} {\bibfnamefont {C.}~\bibnamefont {Bond}}, \bibinfo
  {author} {\bibfnamefont {D.}~\bibnamefont {Brown}}, \bibinfo {author}
  {\bibfnamefont {F.}~\bibnamefont {Br\"uckner}}, \bibinfo {author}
  {\bibfnamefont {L.}~\bibnamefont {Carbone}}, \bibinfo {author} {\bibfnamefont
  {R.}~\bibnamefont {Palmer}}, \ and\ \bibinfo {author} {\bibfnamefont
  {A.}~\bibnamefont {Freise}},\ }\href {\doibase 10.1103/PhysRevD.87.096008}
  {\bibfield  {journal} {\bibinfo  {journal} {Phys. Rev. D}\ }\textbf {\bibinfo
  {volume} {87}},\ \bibinfo {pages} {096008} (\bibinfo {year}
  {2013})}\BibitemShut {NoStop}%
\bibitem [{\citenamefont {Danilishin}\ \emph {et~al.}(2015)\citenamefont
  {Danilishin}, \citenamefont {Gr{\"a}f}, \citenamefont {Leavey}, \citenamefont
  {Hennig}, \citenamefont {Houston}, \citenamefont {Pascucci}, \citenamefont
  {Steinlechner}, \citenamefont {Wright},\ and\ \citenamefont
  {Hild}}]{Danilishin2015}%
  \BibitemOpen
  \bibfield  {author} {\bibinfo {author} {\bibfnamefont {S.~L.}\ \bibnamefont
  {Danilishin}}, \bibinfo {author} {\bibfnamefont {C.}~\bibnamefont
  {Gr{\"a}f}}, \bibinfo {author} {\bibfnamefont {S.~S.}\ \bibnamefont
  {Leavey}}, \bibinfo {author} {\bibfnamefont {J.}~\bibnamefont {Hennig}},
  \bibinfo {author} {\bibfnamefont {E.~A.}\ \bibnamefont {Houston}}, \bibinfo
  {author} {\bibfnamefont {D.}~\bibnamefont {Pascucci}}, \bibinfo {author}
  {\bibfnamefont {S.}~\bibnamefont {Steinlechner}}, \bibinfo {author}
  {\bibfnamefont {J.}~\bibnamefont {Wright}}, \ and\ \bibinfo {author}
  {\bibfnamefont {S.}~\bibnamefont {Hild}},\ }\href
  {http://stacks.iop.org/1367-2630/17/i=4/a=043031} {\bibfield  {journal}
  {\bibinfo  {journal} {New Journal of Physics}\ }\textbf {\bibinfo {volume}
  {17}},\ \bibinfo {pages} {043031} (\bibinfo {year} {2015})}\BibitemShut
  {NoStop}%
\end{thebibliography}%

\end{document}